\documentclass[12pt, a4paper]{article}

\usepackage{amsmath}
\usepackage{amsfonts}
\usepackage{amssymb}
\usepackage{graphicx, rotating}
\usepackage{epstopdf}
\usepackage{epsfig}
\usepackage{latexsym}
\usepackage{graphicx}
\usepackage{color}
\usepackage{amsmath,bm,amssymb}
\usepackage{cite}
\usepackage{slashed}
\usepackage{hyperref}
\hypersetup{colorlinks, citecolor=bluscuro, linkcolor=black, urlcolor=bluscuro}
\definecolor{rossos}{cmyk}{0,1,1,0.55}
\definecolor{bluscuro}{rgb}{0.15, 0.2, .85}
\definecolor{bluchiaro}{cmyk}{1,.3,0.,0.1}

\newcommand{\eq}[1]{Eq.~(\ref{#1})}

\newcommand{\be}{\begin{equation}}
\newcommand{\ee}{\end{equation}}
\newcommand{\bea}{\begin{eqnarray}}
\newcommand{\eea}{\end{eqnarray}}

\def\lra#1{\overset{\text{\scriptsize$\leftrightarrow$}}{#1}}
\def\bma#1{\mbox{\boldmath{$#1$}}}

\begin{document}

\begin{titlepage}
\begin{flushright}
CERN-PH-TH/2013-029
\end{flushright}
\vspace{.3in}

\vspace{1cm}
\begin{center}
{\Large\bf\color{black} Renormalization 
 of dimension-six   operators \\[5mm] 
relevant for the Higgs decays $\bma{h}\rightarrow \bma{\gamma\gamma},
\bma{\gamma Z}$}\\
\bigskip\color{black}
\vspace{1cm}{
{\large  J.~Elias-Mir\'o$^{a,b}$, J.R.~Espinosa$^{a,c,d}$, E.~Masso$^{a,b}$, A.~Pomarol$^{b}$}
\vspace{0.3cm}
} \\[7mm]
{\it {$^a$\, IFAE, Universitat Aut{\`o}noma de Barcelona,
   08193~Bellaterra,~Barcelona}}\\
    {\it {$^b$\, Dept.~de~F\'isica, Universitat Aut{\`o}noma de Barcelona, 08193~Bellaterra,~Barcelona}}\\
{\it $^c$ {ICREA, Instituci\'o Catalana de Recerca i Estudis Avan\c{c}ats, Barcelona, Spain}}\\
{\it $^d$ {CERN, Theory Division, CH--1211 Geneva 23,  Switzerland}}
\end{center}
\bigskip

\vspace{.4cm}

\begin{abstract}
The discovery of the Higgs boson has opened a new  window to test the SM
 through the measurements of its couplings.
Of particular interest is the measured Higgs coupling to photons
which arises in the SM at the one-loop level, and  can then be  
 significantly affected by new physics.
We calculate the   one-loop renormalization of the dimension-six operators
relevant for $h\rightarrow \gamma\gamma, \gamma Z$, which can be potentially  important 
since it could, in principle, give  log-enhanced contributions  from  operator mixing.
We find however that   there is no mixing from any current-current 
operator  that could lead to
this  log-enhanced effect.
We show how  the right choice of  operator basis can  make this  calculation simple.
 We then conclude that  $h\rightarrow \gamma\gamma,  \gamma Z$ can only be affected by RG mixing
from  operators whose Wilson coefficients are expected to be of one-loop size, among them  fermion dipole-moment operators which we have also included.

\end{abstract}
\bigskip

\end{titlepage}

\section{Introduction} 

The discovery   by the LHC \cite{Higgs} of the long-sought Higgs 
boson is a landmark in our quest for understanding the mechanism of electroweak symmetry breaking, which is now open to experimental scrutiny. It is important to measure with precision the Higgs couplings not only to put the  Standard Model (SM) to yet another test, but also because one generically expects deviations from the SM values  in most extensions of the SM, particularly those that address the hierarchy problem. 
Among all experimentally accessible couplings, the Higgs coupling to two photons is particularly interesting.
It has played a central role in the Higgs discovery and, as it arises in the SM at one-loop level, it can be  significantly affected by new physics. Furthermore,  there are tantalizing experimental hints of deviations of the $h\rightarrow \gamma\gamma$ rate 
from SM expectations \cite{Higgs}. 
Another related and interesting Higgs-decay is   
$h\rightarrow \gamma Z$, which is also induced at the one-loop level   in the SM,
and  will be  accessible in the near future.

New-physics effects on SM Higgs decays can be
systematically  studied by means of 
higher-dimensional operators.
This approach is valid whenever 
 the new-physics   mass-scale  $\Lambda$ is  much heavier than the Higgs mass $m_h$, 
a condition that  recent LHC searches seem to suggest.
The purpose of this article is to
 calculate the   renormalization group equations (RGEs)
for the dimension-six operators responsible   for $h\rightarrow \gamma\gamma, \gamma Z$ at the one-loop level.
Our main interest is to look for log-enhanced contributions  coming from  operator mixings.
Particularly interesting are those  contributions  that could arise  from mixings
with operators induced at tree-level by the theory at high-energies.
These can potentially give corrections 
to  the $h\gamma\gamma$ and $h\gamma Z$  couplings of order  $\sim  g_H^2v^2\log (\Lambda/m_h)/(16\pi^2\Lambda^2)$
where  $g_H$ is the coupling of the Higgs to the heavy sector
and $v$ is the Fermi scale.

Recently,  ref.~\cite{GJMT} has argued that 
these type of contributions could in fact be present 
for a general class of models as, for example,  those in ref.~\cite{Giudice:2007fh},
although the result was  based on a  calculation that included  only a partial list of operators and not the complete basis set.
We show however that  such corrections are not present.
The right choice of  operator basis  is  crucial to make the calculation of the anomalous dimensions simple.
We work in a basis 
where the dimension-six operators are classified according to the expected size
of their Wilson coefficients.  We mainly consider two groups:
those operators that can be written as scalar or vector current-current operators
(and  could therefore arise at the tree-level by the interchange of heavy fields),
and the rest,  expected to be induced at the one-loop level.
 By working in this basis,  we  show that none of the  current-current operators affects the running of any  one-loop operator.  This is not a   surprising result, as it is already known to happen in other situations.
For example, the magnetic moment operator responsible for $b\rightarrow s\gamma$ does not  receive log-contributions from   current-current quark operators at the one-loop level \cite{Grinstein:1990tj}.

We  also show how to reconcile our conclusion with the results of \cite{GJMT} by completing the calculation done in the basis used in that analysis.
Furthermore, we use the results of ref.~\cite{GJMT} to calculate the complete leading-log corrections
to the operators responsible for  $h\rightarrow \gamma\gamma$ and $h\rightarrow \gamma Z$.
 This is only affected by  Wilson coefficients  of one-loop operators, 
  and therefore  these effects are not expected to be very large.
 Finally, we also extend the calculation to include mixing with  fermion dipole-moment operators.

\section{Dimension-six operator basis} 
\label{Dimension-six Operator Basis}

Whenever  the mass-scale of new physics   
$\Lambda$ is larger than the relevant energy-scale  involved in a SM process, 
we can parametrize all new-physics effects by 
higher-dimensional local operators made from   an expansion  in
\begin{equation}
\frac{D_\mu}{\Lambda}\ ,\  \frac{g_H H}{\Lambda}\ ,\  \frac{g_{f_{L,R}}  f_{L,R}}{\Lambda^{3/2}}\ ,\ \frac{gF_{\mu\nu}}{\Lambda^2}\, .
\label{expansion}
\end{equation}
We denote by $D_\mu$  the covariant derivatives,  $g_H$ and $g_{f_{L,R}}$     respectively  account for the couplings of the Higgs-doublet field $H$ and SM fermion $f_{L,R}$ to the new heavy sector, while  $g$ and $F_{\mu\nu}$   are  the SM gauge couplings and field-strengths.
 At  leading order in this expansion,  and assuming lepton number is conserved,
the dominant operators are  of dimension six.
It is very important to choose the right 
set of independent dimension-six operators  that defines a complete basis.
A suitable basis is one which can capture  in a  simple way  
 the impact of different   new-physics scenarios.
Since  usually a given  new-physics scenario only generates a sub-class of  operators,
it is  convenient to choose a basis that does not mix these sub-classes, at least for the most interesting scenarios.
Another important requirement  for the basis
is that it should not mix operators  whose coefficients are naturally expected to have very different sizes.
For example,  tree-level  operators,  that can be induced in weakly-coupled renormalizable theories, should be kept separate from one-loop induced ones.  As already said, this is also important since, at the one-loop level, 
it is   frequently  found that tree-level induced operators  do not  contribute
to the RG flow of one-loop induced ones. 

Let us start  considering only   operators made of SM bosons. 
These can be induced from  integrating out  heavy states in  "universal theories", those
whose fields only couple to the bosonic sector of the SM.
(A generalization including  SM fermions will be given later.)
The  appropriate basis  was defined in
ref.~\cite{Giudice:2007fh} and in it we can broadly distinguish three classes of operators.
The first two classes consist of  operators that can in principle be generated at  tree-level when integrating out 
heavy  states with spin $\leq 1$  
under the assumption of  minimal-coupling as defined in ref.~\cite{Giudice:2007fh} (or, alternatively,
induced at tree-level from weakly-coupled  renormalizable theories).
 The operators of the first class are those that involve extra powers of Higgs fields, and  are expected to be  suppressed by  $g_H^2/\Lambda^2$.
Since  $g_H$ can be as large as  $\sim 4\pi$,  the effects of these operators
can dominate over the rest. 
The operators of the second class involve extra (covariant) derivatives 
or gauge-field strengths and, according to \eq{expansion},  are generically suppressed by $1/\Lambda^2$.
 Finally,  in the  third class, we consider operators that,
 in minimally-coupled theories, 
  can only be induced  at the one-loop level. These operators are  
 expected to be suppressed by   $g_H^2/(16\pi^2\Lambda^2)$,
 although they could  be  further suppressed by an extra factor $g^2/g_H^2$ if the 
 external fields are gauge bosons.

We can then classify the dimension-six operators as 
\begin{equation}
{\cal L}_{6} =\sum_{i_1}g^2_H\frac{c_{i_1}}{\Lambda^2}{\cal O}_{i_1}+\sum_{i_2}\frac{c_{i_2}}{\Lambda^2}{\cal O}_{i_2}+\sum_{i_3}\frac{\kappa_{i_3}}{\Lambda^2}{\cal O}_{i_3}\, ,
\label{6dim}
\end{equation}
where   for notational convenience we introduce for the third type of operators the one-loop suppressed coefficients 
\be
\kappa_{i_3}\equiv \frac{g_H^2}{16\pi^2} c_{i_3}\, .
\ee
All coefficients $c_i$ are of order  $c_i\sim O(1)\times  f(g/g_H,...)\lesssim O(1)$, with $f(g/g_H,...)$ a function that depends only on  ratios of couplings  and is not expected to be larger than order one.  
In the first  class of operators, ${\cal O}_{i_1}$, suppressed by $g^2_H/\Lambda^2$, we have~\footnote{In  ${\cal O}_6$ we have replaced a factor $g_H^2$ by  a factor $\lambda$,  the Higgs self-coupling, 
as this is what appears in  theories in which the Higgs is protected by a symmetry. 
Similarly, for operators involving  $\bar f_Lf_R H$ we include a Yukawa coupling,  as in (\ref{oy}). }
\begin{equation}
{\cal O}_H=\frac{1}{2}(\partial^\mu |H|^2)^2\ \  ,\ \ \
{\cal O}_T=
\frac{1}{2}\left (H^\dagger {\lra{D}_\mu} H\right)^2
 \ \ ,\ \ \
{\cal O}_r=|H|^2 |D_\mu H|^2\ \ ,\ \ \
{\cal O}_6=\lambda |H|^6\, .
\label{first6dim}
\end{equation}
Here we have defined $H^\dagger {\lra { D_\mu}} H\equiv H^\dagger D_\mu H - (D_\mu H)^\dagger H $, with
$D_\mu H = \partial_\mu H -i g\sigma^a W^a_\mu H/2 - i g' B_\mu H/2$, the standard covariant derivative (our Higgs doublet, 
$H=(G^+,(h+iG^0)/\sqrt{2})^T$, has hypercharge $Y=1/2$). 
Finally, $\lambda$
is the  Higgs quartic coupling in the SM potential, $V= m^2 |H|^2+\lambda |H|^4$.
By means of the redefinition $H\rightarrow H[1-c_rg^2_H|H|^2/(2\Lambda^2)]$
 we could trade the operator ${\cal O}_r$  with \cite{Giudice:2007fh}
\begin{equation}
{\cal O}_y   =|H|^2 \left[ y_u  {\bar Q}_L \widetilde{H}  u_R +y_d  {\bar Q}_L  H d_R + y_l {\bar L}_L H l_R
\right] \, ,
\label{oy}
\end{equation}
where sum over all families is understood, and $\widetilde{H}=i\sigma^2 H^*$. Here $y_f$ are Yukawa couplings, normalized as usual, with $m_f=y_f v/\sqrt{2}$ and $v=\langle h\rangle=246$ GeV.

In the  second class of operators, ${\cal O}_{i_2}$, suppressed by $1/\Lambda^2$,  we have~\footnote{
The  operator ${\cal O}_{4K}=|D_\mu^2  H|^2$ can be
eliminated  by a field redefinition of $H$. See Appendix for details.}
\begin{eqnarray}
&&{\cal O}_W=\frac{ig}{2}\left( H^\dagger  \sigma^a \lra {D^\mu} H \right )D^\nu  W_{\mu \nu}^a\ \ ,\ \ \
{\cal O}_B=\frac{ig'}{2}\left( H^\dagger  \lra {D^\mu} H \right )\partial^\nu  B_{\mu \nu}\, ,\nonumber\\
&&{\cal O}_{2W}=-\frac{1}{2}  ( D^\mu  W_{\mu \nu}^a)^2\ \ ,\ \ \
{\cal O}_{2B}=-\frac{1}{2}( \partial^\mu  B_{\mu \nu})^2\ \ ,\ \ \
{\cal O}_{2G}=-\frac{1}{2}  ( D^\mu  G_{\mu \nu}^a)^2\, .
\label{second6dim}
\end{eqnarray}
The easiest way to see that the operators of  \eq{first6dim} and  \eq{second6dim}
can be generated at tree-level  is to realize that they 
 can   be written as  products of vector and scalar currents \cite{Giudice:2007fh,LRV}. For example, ${\cal O}_T=(1/2) {J_H}^\mu {J_H}_\mu $, where ${J_H}^\mu=H^\dagger {\lra { D^\mu}} H$,   could arise from integrating out a massive vector.
We will refer to the operators (\ref{first6dim}) and  (\ref{second6dim}) as "current-current" or "tree-level" operators.

In the third class of operators, ${\cal O}_{i_3}$, suppressed by an extra  loop factor, we have the CP-even operators
\begin{eqnarray}
&&{\cal O}_{BB}={g}^{\prime 2} |H|^2 B_{\mu\nu}B^{\mu\nu} \ \ ,\ \ \
{\cal O}_{GG}=g_s^2 |H|^2 G_{\mu\nu}^a G^{a\mu\nu}\label{third6dim1}\, , \\
&&{\cal O}_{HW}=i g(D^\mu H)^\dagger\sigma^a(D^\nu H)W^a_{\mu\nu}\ \ ,\ \ \
{\cal O}_{HB}=i g'(D^\mu H)^\dagger(D^\nu H)B_{\mu\nu}\label{third6dim2}\, ,
\\
&&{\cal O}_{3W}=g\epsilon_{abc}W^{a\, \nu}_{\mu}W^{b}_{\nu\rho}W^{c\, \rho\mu}\ \ ,\ \ \
{\cal O}_{3G}=g_s f_{abc}G^{a\, \nu}_{\mu}G^{b}_{\nu\rho}G^{c\, \rho\mu}\, ,
\label{third6dim3}
\end{eqnarray}
and the CP-odd operators
\begin{eqnarray}
\label{third6dimCP1}
&&{\cal O}_{B\widetilde B}={g}^{\prime 2} |H|^2 B_{\mu\nu}\widetilde B^{\mu\nu} \ \ ,\ \ \
{\cal O}_{G\widetilde G}=g_s^2 |H|^2 G_{\mu\nu}^a \widetilde G^{a\mu\nu}\, ,\\
\label{third6dimCP2}
&&{\cal O}_{H\widetilde W}=g(D^\mu H)^\dagger\sigma^a(D^\nu H)\widetilde W^a_{\mu\nu}\ \ ,\ \ \
{\cal O}_{H\widetilde B}=g'(D^\mu H)^\dagger(D^\nu H)\widetilde B_{\mu\nu}\, ,
\\
&&{\cal O}_{3\widetilde W}=g\epsilon_{abc}\widetilde W^{a\, \nu}_{\mu}W^{b}_{\nu\rho}W^{c\, \rho\mu}\ \ ,\ \ \
{\cal O}_{3\widetilde G}=g_s f_{abc}\widetilde G^{a\, \nu}_{\mu}G^{b}_{\nu\rho}G^{c\, \rho\mu}\, ,
\label{third6dimCP3}
\end{eqnarray}
where $\widetilde F^{\mu\nu}=\epsilon^{\mu\nu\rho\sigma}F_{\rho\sigma}/2$.
We will refer to these operators as "one-loop suppressed"  operators.

We emphasize again that 
the above classification is useful even when one is not working under the minimally-coupled assumption of 
ref.~\cite{Giudice:2007fh}. When studying the RGEs  of these operators,
we will find that, at  leading order,
current-current operators do not affect the RG running 
of   one-loop suppressed  operators (irrespective of their UV origin).
 Furthermore, the above classification  can  also be useful to parametrize  the effects of strongly-coupled models.
In particular, if the Higgs is part of the composite meson states,
taking $g_H\sim 4\pi $ gives the correct power counting  for  strongly-coupled theories with no small parameters.
One finds in this case  that   operators of the first class are the most relevant,
  while   operators of the second and third class have  the same $1/\Lambda^2$ suppression.
Also  the basis is suited for characterizing holographic descriptions of strongly-coupled models \cite{Giudice:2007fh}.    
In this case $g_H \sim 4\pi/\sqrt{N}$, where $N$ plays the role of the number of colors of the strong-interaction,
and then operators of the  first and second class   are less suppressed than
operators of the third class.

\section{
Non-renormalization  of  $\bma{h\rightarrow \gamma\gamma,\gamma Z}$ 
from  current-\\ current  operators}
\label{section3}

The operator basis introduced in the previous section is particularly well-suited
to describe new-physics contributions to  $h\rightarrow \gamma\gamma$, which
come only from two operators: the CP-even ${\cal O}_{BB}$ and the CP-odd ${\cal O}_{B\widetilde B}$. On the other hand, 
 $h\rightarrow \gamma Z$ comes (on-shell) from ${\cal O}_{BB}$, ${\cal O}_{HB}$,  ${\cal O}_{HW}$
 and their CP-odd counterparts. The relevant Lagrangian terms for such decays are
\bea
\delta {\cal L}_{\gamma\gamma} &=& \frac{e^2}{2\Lambda^2}
\Big[\kappa_{\gamma\gamma}\, h^2 F_{\mu\nu}F^{\mu\nu}
+ \kappa_{\gamma\widetilde\gamma}\, h^2 F_{\mu\nu}\widetilde F^{\mu\nu}\Big]\, ,\nonumber\\
\delta {\cal L}_{\gamma Z} &=& \frac{e\, G}{2\Lambda^2}
 \Big[\kappa_{\gamma Z}\, h^2 F_{\mu\nu}Z^{\mu\nu}+
 \kappa_{\gamma \widetilde Z}\,  h^2 F_{\mu\nu}\widetilde Z^{\mu\nu}\Big]\, ,
\label{gZg}
\eea
 where  $e=gg'/G$ and $G^2= g^2+{g'}^2$. 
 The photon field, $A_\mu=c_{w} B_\mu+s_w  W^3_\mu$, has field-strength $F_{\mu\nu}$, while 
$Z_\mu=c_w  W^3_\mu-s_w B_\mu$ has field-strength $Z_{\mu\nu}$, where we use $s_w\equiv \sin\theta_w=g'/G$ and $c_w\equiv \cos\theta_w=g/G$. We have
\bea
\label{cggcgz}
 \kappa_{\gamma\gamma}=\kappa_{BB}\ ,\quad &&
 \kappa_{\gamma Z}= \frac{1}{4}(\kappa_{HB}-\kappa_{HW})-2s_w^2 \kappa_{BB}\, ,\nonumber\\
  \kappa_{\gamma\widetilde\gamma}=\kappa_{B\widetilde B}\ ,\quad&&
  \kappa_{\gamma \widetilde Z}=\frac{1}{4}(\kappa_{H\widetilde B}-\kappa_{H\widetilde W})-2s_w^2 \kappa_{B\widetilde B}\, .
\eea
The Wilson coefficients of these   dimension-six operators are generated at the scale $\Lambda$, at which the heavy new physics is integrated out, and they  should be renormalized down to the Higgs mass, at which they  are  measured
in  Higgs decays.
Let us focus for simplicity on $\kappa_{\gamma\gamma}$, as similar considerations will be applicable to $\kappa_{\gamma\widetilde\gamma}, \kappa_{\gamma Z}, \kappa_{\gamma \widetilde Z}$. At one-loop leading-log order one has, running from $\Lambda$ to the Higgs mass $m_h$:
\be
\kappa_{\gamma\gamma}(m_h) = \kappa_{\gamma\gamma}(\Lambda) - \gamma_{\gamma\gamma} \log \frac{\Lambda}{m_h}\ .
\ee
 Here, $\gamma_{\gamma\gamma}=d \kappa_{\gamma\gamma}/d\log \mu$, with $\mu$ the energy scale, is the one-loop anomalous dimension for $\kappa_{\gamma\gamma}$.
In principle, $\gamma_{\gamma\gamma}$  can depend on the Wilson coefficients of any  dimension-six operator in Eq.~(\ref{6dim}). 
A particularly interesting case would be if 
the RGEs
 were to mix the tree-level operators into the RG evolution of one-loop suppressed operators, such as ${\cal O}_{BB}$. In that case  we   would expect $\gamma_{\gamma\gamma}\sim g^2_H/(16\pi^2)$  from  mixings with the operators  of \eq{first6dim},
 or $\gamma_{\gamma\gamma}\sim g^2/(16\pi^2)$ from  mixings with (\ref{second6dim}).
Such loop effect could   give a sizeable contribution to $\kappa_{\gamma\gamma}(m_h)$,  logarithmically enhanced by a 
factor $\log{\Lambda}/{m_h}$. The initial   value  $\kappa_{\gamma\gamma}(\Lambda)$,   expected to be one-loop suppressed, would  then be subleading.

Remarkably, and this is our main result,   there is no mixing from  tree-level  operators (\ref{first6dim})-(\ref{second6dim}) 
to  one-loop suppressed  operators (\ref{third6dim1})-(\ref{third6dimCP3}), at least  at the one-loop level.
This can be easily  shown for the  renormalization of $\kappa_{\gamma\gamma}$. 
The argument goes as follows.
Let us  first  consider the effects of the first-class operators, \eq{first6dim}.
Since  these  operators  have four or more $H$,
their contribution  to the renormalization of $\kappa_{\gamma\gamma}$  can only arise from a  loop of the
electrically-charged $G^\pm$ with  at least one photon  attached to the loop. However,
\begin{itemize}
\item{${\cal O}_6$} has too many Higgs legs to contribute. 
\item{${\cal O}_H$ is simply $\partial_\mu(h^2+G_0^2+2G^+G^-)
\partial^\mu(h^2+G_0^2+2G^+G^-)/8$}  and this momentum structure implies that a $G^\pm$ loop can only give  a contribution  $\propto \partial_\mu h^2$, which  is not the Higgs momentum structure of  \eq{gZg}.
\item{${\cal O}_T$} does not contain a vertex $h^2G^+G^-$.
\item{${\cal O}_r$}  can be traded with ${\cal O}_y$, which clearly can only give one-loop contributions to operators $\propto |H|^2H$, so it only contributes to the RGE of itself and ${\cal O}_6$.
\end{itemize}
We conclude that there is no contribution from these operators to the RGE of $\kappa_{\gamma\gamma}$.
To  generalise the proof  that no  operator in  (\ref{first6dim}) contributes   to the one-loop anomalous-dimension 
of  any  operator  in (\ref{third6dim1})-(\ref{third6dim3})~\footnote{Obviously, their contribution to the CP-odd operators (\ref{third6dimCP1})-(\ref{third6dimCP3}) is zero as the SM  gauge-boson couplings conserve CP.},
 we have calculated  explicitly the one-loop  operator-mixing.
We find that  the only  operators involving two Higgs  and  gauge bosons
that can be affected by (\ref{first6dim}) 
are the tree-level operators  (\ref{second6dim}). 
The result is  given in Section~4.

For the operators  of \eq{second6dim},
proving the absence of one-loop contributions  to the anomalous dimension of 
(\ref{third6dim1})-(\ref{third6dim3})  is  even simpler.
By means of  field redefinitions, as those given in the Appendix, or, equivalently,
by using the equations of motion~\footnote{That is, $2D^\nu  W_{\mu \nu}^a
=igH^\dagger  \sigma^a \lra {D_\mu} H +g\bar f_L\sigma^a\gamma_\mu  f_L$
and 
$\partial^\nu  B_{\mu \nu}=ig' H^\dagger  \lra {D_\mu} H/2+g'
Y_{L}^f\bar f_L \gamma_\mu f_L+g'Y_{R}^f\bar f_R\gamma_\mu  f_R$,
where $Y_{L,R}^f$ are the fermion hypercharges and a sum over fermions is understood.}, 
we can trade the  operators (\ref{second6dim}) with
operators of \eq{first6dim}, four-fermion operators and  operators of the type
\bea
{\cal O}_{R}^f&=&
(i \, H^\dagger {\lra { D_\mu}} H)( \bar f_R\gamma^\mu f_R),\
\nonumber\\
{\cal O}_{L}^f&=&
(i \, H^\dagger {\lra { D_\mu}} H)( \bar f_L\gamma^\mu f_L),\ 
\nonumber\\
{\cal O}_{L}^{f\, (3)}&=&
(i \, H^\dagger \sigma^a {\lra { D_\mu}} H)( \bar f_L\gamma^\mu\sigma^a f_L)\, .
\label{first6dimF}
\eea
Now, four-fermion operators contain too many fermion legs to  contribute to operators made only of SM bosons. Concerning
 the operators of \eq{first6dimF}, after  closing the fermion legs in a loop,  it is clear that they can only  give contributions to operators with the Higgs structure $H^\dagger {\lra { D_\mu}} H$ or
$H^\dagger \sigma^a{\lra { D_\mu}} H$, corresponding to the  tree-level operators (\ref{second6dim}).
This completes the proof that  no current-current  operator contributes to the running  of any one-loop suppressed operator.

The calculation above could have also been done in 
other operator bases. To keep  the calculation simple, it is crucial  to work in bases that do not  mix current-current operators  with one-loop suppressed ones.
This is guaranteed if we change basis  by means of SM-field  redefinitions, as shown in  the Appendix. We can make use of these field-redefinitions to work in bases that contain only 3 operators made of bosons, the rest consisting of operators involving fermions, such as those in  \eq{oy}, \eq{first6dimF} or 4-fermion operators.
There are different options in choosing these 3 operators; what is   
physically relevant are the 3 (shift-invariant) combinations of  coefficients in \eq{physcoef}.
This freedom  can be used to select the set of 3 operators
most convenient to prove, in the simplest way, that
their contribution  to the running
of $\kappa_{\gamma\gamma}$ and $\kappa_{Z\gamma}$ is zero at the one-loop level.
For example,  we could have chosen ${\cal O}_{2B}$ instead of  ${\cal O}_{T}$: since ${\cal O}_{2B}$ only affects the propagator of the  neutral state  $B^\mu$,
one can easily see that it cannot 
contribute to the $h\gamma\gamma$ or $h\gamma Z$ coupling.

Let us finally mention that there is an alternative way to see that the running of $\kappa_{\gamma\gamma}$
is not affected at the one-loop level by  tree-level operators.
This corresponds to showing that  any  heavy charged state of mass $M$,  coupled to photons only through the covariant
derivative, gives at the one-loop level a contribution 
to the  effective $h\gamma\gamma$ coupling that does not contain terms like $\log M/m_h$
(which in the effective theory below $M$ are interpreted as the running from $M$ to $m_h$).
We can easily show the absence of  such logarithms
by working in the limit  $M\gg m_h$ where
 we can use   low-energy theorems  \cite{Shifman:1979eb}
 to relate the $h\gamma\gamma$ coupling    to the 
two-point function of the photon. At the one-loop level we have 
\be
\frac{\kappa_{\gamma\gamma}(\mu)}{\Lambda^2}=-\frac{1}{4v}\left. \frac{\partial }{\partial  h}\,  \frac{1}{e^{2}_{\rm eff}(\mu, h)}\right|_{h=v}\, , 
\label{hgg2f}
\ee
where $e_{\rm eff}(\mu, h)$  is  the effective electric coupling  calculated in a nonzero
Higgs background:
\be
\frac{1}{e^{2}_{\rm eff}(\mu, h)}=\frac{1}{e^{2}(\Lambda_{\rm UV})}+
\frac{b_a}{16\pi^2} \log\frac{M(h)}{\Lambda_{\rm UV}}+
\frac{b_b}{16\pi^2} \log\frac{\mu}{M(h)}\, ,
\label{betag}
\ee
with $b_{a,b}$ being respectively the beta-function of the gauge coupling above and below $M(h)$, the   mass of the heavy state  in the Higgs background.
From \eq{hgg2f} and \eq{betag}  we have
\be
\gamma_{\gamma\gamma}=\frac{\Lambda^2}{16\pi^2}
\left.  \frac{d}{d\log\mu}\left[\frac{(b_b-b_a)}{4vM(h)}\frac{\partial M(h)}{\partial h}\right]\right|_{h=v}=0\, ,
\ee
due to the fact   that  $b_{a,b}$ are independent of $\mu$
at the one-loop level. Simply put, a heavy charged particle
with mass $M$  contributes to the running of the photon two-point function through a
loop which only contains that particle itself, and therefore
no log-terms involving the light-state masses  are possible.

\section{The importance of the choice of basis}

The relevance of the possible contributions from   tree-level operators
to the one-loop RGE of $\kappa_{\gamma\gamma}$ and $\kappa_{\gamma Z}$
has been highlighted recently in ref.~\cite{GJMT}. In fact, that analysis
claims that such important effect could actually occur, in contradiction
with the results presented in the previous section. 
In this section we show how this contradiction is resolved.

The analysis in ref.~\cite{GJMT}, GJMT in what follows, focuses on a subset of dimension-six operators, chosen to be ${\cal O}_{BB}$
and the two operators 
\be
{\cal O}_{WB}=g g' (H^\dagger\sigma^a H) W^a_{\mu\nu}B^{\mu\nu} \ ,\quad
{\cal O}_{WW}=g^2|H|^2 W^a_{\mu\nu}{W^a}^{\mu\nu} \ ,
\ee
which are not included in the basis we have used. The relation to our basis follows from the two operator identities:
\begin{eqnarray}
&&
{\cal O}_B={\cal O}_{HB}+\frac{1}{4}{\cal O}_{WB}+\frac{1}{4}{\cal O}_{BB}\ ,\label{OpId1}\\
&&{\cal O}_W={\cal O}_{HW}+\frac{1}{4}{\cal O}_{WW}+\frac{1}{4}{\cal O}_{WB}\ ,\label{OpId2}
\end{eqnarray}
which allow us to remove ${\cal O}_{WW}$ and ${\cal O}_{WB}$
in favor of ${\cal O}_B$ and ${\cal O}_W$. The two operators ${\cal O}_{HW}$ and ${\cal O}_{HB}$ were also mentioned in  ref.~\cite{GJMT}, although their effect was not included in the analysis.
To understand the issues involved it will be sufficient to limit the operator basis to five operators, with the two bases used being
\bea
\label{B1}
B_1 &=& \{{\cal O}_{BB},{\cal O}_{B},{\cal O}_W, {\cal O}_{HW}, {\cal O}_{HB}\}\ ,\quad (\mathrm{this}\ \mathrm{work)}\\
B_2 &=&\{{\cal O}_{BB}, {\cal O}_{WW},{\cal O}_{WB},{\cal O}_{HW}, {\cal O}_{HB}\}\ ,\quad {\mathrm{(GJMT)}}.
\label{B2}
\eea
In relating both bases we will use primed Wilson coefficients for the GJMT
basis
\be
{\cal L}_6=\sum_i \frac{c'_i}{\Lambda^2}{\cal O}_i\, ,
\ee
and the dictionary to translate between $B_1$ and $B_2$ is:
\bea
\kappa_{HW}&=&c'_{HW}-4c'_{WW}\ , \nonumber\\
\kappa_{HB}&=&c'_{HB}+4(c'_{WW}-c'_{WB})\ , \nonumber\\
\kappa_{BB}&=&c'_{BB}+c'_{WW}-c'_{WB}\ , \nonumber\\
 c_W&=& 4 c'_{WW}\ , \nonumber\\
 c_B&=&4(c'_{WB}-c'_{WW})\, .
\eea
From these relations we can directly write the expressions for $\kappa_{\gamma\gamma}$ and $\kappa_{\gamma Z}$ going from (\ref{cggcgz}) to the GJMT basis:
\bea
\label{cgg2}
\kappa_{\gamma\gamma}&=&c'_{BB}+c'_{WW}-c'_{WB}\ ,\nonumber\\
\kappa_{\gamma Z}&=&2c_w^2 c'_{WW}-2s_w^2 c'_{BB} -(c_w^2-s_w^2)c'_{WB}+\frac{1}{4}(c'_{HB}-c'_{HW})\ .
\eea

 Let us first note  that the operator identities (\ref{OpId1}) and (\ref{OpId2})
show that  two operators of the GJMT basis, ${\cal O}_{WW}$ and ${\cal O}_{WB}$, are a mixture of
tree-level operators and one-loop suppressed  ones of   basis $B_1$.
This has   the following drawback.
Let us suppose that 
the operator 
${\cal O}_{W}$ is generated, for example,
by integrating out a  heavy SU(2)-triplet gauge boson
(see {\it e.g.} \cite{LRV}).
This operator can  be
 written in the GJMT basis by using the  identity (\ref{OpId2}), but then
the  coefficients of the operators ${\cal O}_{WW}$, ${\cal O}_{WB}$ and ${\cal O}_{HW}$ generated in this way will all be correlated. 
In this particular example, we will have $c_{WW}'=c_{WB}'=c_{HW}'/4$.
This is telling  us  that when   using the GJMT basis 
 to study the physical impact of  this scenario
 we must include the effects of all  operators, 
and not only a partial list of them, as done in  ref.~\cite{GJMT}.
Otherwise,  one can miss  contributions of the same size that could lead to cancellations.
The same argument goes through for scenarios generating the tree-level operator ${\cal O}_B$. 
In general, the correlation of the coefficients in the GJMT basis  is  explicitly  shown in the reversed dictionary:
\bea
c'_{WW}&=&\frac{1}{4}c_W \ , \nonumber\\
c'_{WB}&=&\frac{1}{4}(c_B+c_W) \ , \nonumber\\
c'_{BB}&=&\frac{1}{4}c_B+\kappa_{BB}\ , \nonumber\\
c'_{HW}&=&c_W+\kappa_{HW}\ , \nonumber\\
c'_{HB}&=&c_B+\kappa_{HB} \ .
\label{invdict}
\eea
Obviously,  physics does not depend on what basis is used, 
which is a matter of choice, as long as  the full calculation is done in both bases.
Reducing, however, the calculations to a few operators in a given basis can be dangerous
as this can leave out important effects.
This is   especially   true in bases whose operators   are a mixture of operators with  Wilson coefficients of different sizes.
For this reason the basis $B_1$ is preferable to $B_2$.

To explicitly show  how this correlation between  Wilson coefficients
can  lead to cancellations in the final result, 
let us consider a  particularly simple  example:
 the calculation of the radiative corrections 
to the operators ${\cal O}_{WW}$, ${\cal O}_{BB}$
and ${\cal O}_{WB}$ proportional to $\lambda$.
This is partly given   in the analysis of \cite{GJMT},   apparently 
showing a  one-loop mixing from tree-level operators to one-loop suppressed ones.
As obtained in \cite{GJMT}, the $\lambda$-dependent piece of the 
anomalous-dimension matrix for $c'_{BB}, c'_{WW}, c'_{WB}$ is given by
\be
\frac{d}{d\log \mu}\left[\begin{array}{c}
c'_{BB}\\
c'_{WW}\\
c'_{WB}
\end{array}\right]
=\frac{1}{16\pi^2}
\left(\begin{array}{ccc}
12\lambda &0&0\\
0&12\lambda &0\\
0&0&4\lambda 
\end{array}\right)
\left[\begin{array}{c}
c'_{BB}\\
c'_{WW}\\
c'_{WB}
\end{array}\right]+...\, .
\label{betasGJMT}
\ee 
From (\ref{cgg2}), one obtains the RGE 
\be
\label{betagg}
\gamma_{\gamma\gamma}=
\frac{d \kappa_{\gamma\gamma}}{d\log \mu}=\frac{4\lambda}{16\pi^2}
(3 \kappa_{\gamma\gamma}+2 c'_{WB})+...\, , 
\ee
showing explicitly that  the coefficient $c'_{WB}$, which can be of tree-level size in the GJMT basis [see  (\ref{invdict})], affects the running of the one-loop suppressed $\kappa_{\gamma\gamma}$. 
This apparent contradiction with our previous result  is, as expected,  resolved by adding 
the effect of the operators ${\cal O}_{HW}$ and ${\cal O}_{HB}$ in the renormalization of $\kappa_{\gamma\gamma}$. 
We obtain the ($\lambda$-dependent) contributions
\be
\frac{d c'_{BB}}{d\log\mu} =-\frac{3\lambda}{16\pi^2}  c'_{HB}\ ,\quad
\frac{d c'_{WW}}{d\log\mu} =-\frac{3\lambda}{16\pi^2}  c'_{HW}\ ,\quad
\frac{d c'_{WB}}{d\log\mu} =-\frac{\lambda}{16\pi^2}(c'_{HB}+c'_{HW})\ ,
\ee
which change the RGE (\ref{betagg}) into
\be
\gamma_{\gamma\gamma}= \frac{2\lambda}{16\pi^2}
(6\kappa_{\gamma\gamma}+4 c'_{WB}- c'_{HB}- c'_{HW})\ .
\ee
These additional contributions   eliminate the  possibly sizeable tree-level correction from $c'_{WB}$. 
Indeed, using (\ref{invdict}), we explicitly
see that the contributions proportional to  $c_W$ and $c_B$ cancel out, giving
\be
\gamma_{\gamma\gamma}= \frac{2\lambda}{16\pi^2}
\Big(6 \kappa_{\gamma\gamma}- \kappa_{HB}-\kappa_{HW}\Big)\ ,
\ee
leaving behind just  corrections from one-loop suppressed operators.
{This is not an accident: this cancellation was expected from our discussion in the previous section. Beyond the $\lambda$-dependent terms we have examined, the same cancellation will necessarily occur for the rest of the potentially sizeable contributions to $\gamma_{\gamma\gamma}$ identified in \cite{GJMT}.

\section{Renormalization group equation for $\bma{\kappa_{\gamma\gamma}}$ and $\bma{ \kappa_{\gamma\widetilde\gamma}}$ 
\label{sec:rencgg}}

In this section we use the results of ref.~\cite{GJMT}, combined with our results in section \ref{section3}, to obtain $\gamma_{\gamma\gamma}$.
Let us write the RGEs  for the Wilson coefficients in basis $B_2$ in a compact way as
\be
16\pi^2\frac{d c'_i}{d\log\mu}=\sum_{j=1}^5 b'_{i,j}c'_j\ .
\ee
The $b'_{i,j}$ is a $5\times 5$  anomalous-dimension matrix
of which the $3\times 3$ submatrix corresponding to $i,j=1-3$ (that is, $c'_{BB}$, $c'_{WW}$, $c'_{WB}$) was
calculated in \cite{GJMT}, while the rest is unknown.
From  $\kappa_{\gamma\gamma}=\sum_{i=1}^5\zeta_i c'_i$ where  $\zeta_i=(1,1,-1,0,0)$, we have
\be
16\pi^2\gamma_{\gamma\gamma}= \sum_{i,j=1}^5\zeta_i b'_{i,j}c'_j\, .
\ee 
Using Eq.~(\ref{invdict}), we can translate this
anomalous dimension to our basis. We get
\bea
\label{betagg1}
16\pi^2\gamma_{\gamma\gamma}&=&\sum_{i=1}^5 \zeta_i ( b'_{i,BB}\kappa_{BB}+b'_{i,HW}\kappa_{HW}+b'_{i,HB}\kappa_{HB})\\
&
+&\frac{1}{4}c_B\sum_{i=1}^5 \zeta_i (b'_{i,WB}+b'_{i,BB}+4b'_{i,HB})
+\frac{1}{4}c_W\sum_{i=1}^5 \zeta_i (b'_{i,WW}+b'_{i,WB}+4b'_{i,HW})\ .\nonumber
\eea
From our discussion in Section~\ref{Dimension-six Operator Basis}, we know that the tree-level coefficients $c_B$ and $c_W$ do not appear in this RGE. 
This means that the two last terms of \eq{betagg1} must be zero, allowing 
 us  to extract the sum of the unknown coefficients $b'_{i,HB}$
and $b'_{i,HW}$ in terms of  coefficients calculated in ref.~\cite{GJMT}:
\be
\sum_{i=1}^5\zeta_i b'_{i,HB}=-\frac{1}{4}\sum_{i=1}^5 \zeta_i(b'_{i,WB}+b'_{i,BB})\ ,\quad
\sum_{i=1}^5\zeta_i b'_{i,HW}=-\frac{1}{4}\sum_{i=1}^5 \zeta_i(b'_{i,WW}+b'_{i,WB})\, .
\label{inter}
\ee
Notice that $\zeta_4=\zeta_5=0$ is crucial to allow us to restrict the sums in the right-hand-side to terms that were already calculated 
in \cite{GJMT}. Plugging the terms (\ref{inter}) back in (\ref{betagg1}), one gets
\be
16\pi^2\gamma_{\gamma\gamma}=\sum_{i=1}^5 \zeta_i
\left[b'_{i,BB} \kappa_{BB}-\frac{1}{4}(b'_{i,WB}+b'_{i,WW})\kappa_{HW}-\frac{1}{4}(b'_{i,BB}+b'_{i,WB})\kappa_{HB}\right]\ .
\ee
Using the coefficients $b'_{i,WW}, b'_{i,WB}$ and $b'_{i,BB}$ from \cite{GJMT}, one arrives at
\be
16\pi^2\gamma_{\gamma\gamma}=\left[6y_t^2-\frac{3}{2}(3g^2+{g'}^2)+12\lambda\right]\kappa_{BB}
+
\left[\frac{3}{2}g^2-2\lambda\right](\kappa_{HW}+\kappa_{HB})\ .
\label{final}
\ee
This expression gives the  one-loop leading-log correction to $\kappa_{\gamma\gamma}(m_h)$.
For the  resummation of the log terms  we would need the   full anomalous-dimension matrix. Nevertheless,
this is not needed for $\Lambda\sim $ TeV since the log-terms are not very large.

The size of the contributions of  \eq{final} to $\kappa_{\gamma\gamma}(m_h)$
is expected to be of two-loop order in  minimally-coupled theories. 
Therefore, we have to keep in mind that  the tree-level operators of  \eq{first6dim}, possibly entering in the RGE of $\kappa_{\gamma\gamma}$
 at the two-loop level,  could give corrections of the same order.
For strongly-coupled theories in which $g_H\sim 4\pi$,
we could have $\kappa_i\sim O(1)$, and the corrections from \eq{final}   to $h\rightarrow \gamma\gamma$ could be of one-loop size. 
Of course, in principle,   the initial values    $\kappa_i(\Lambda)$  will give,
 as    
\eq{cggcgz} shows,
 the dominant contribution to
  $h\rightarrow \gamma\gamma, \gamma Z$ and not \eq{final}.
Nevertheless,  it could well be the case that $|\kappa_{BB}(\Lambda)|\ll 1$  and  $|\kappa_{HB}(\Lambda)- \kappa_{HW}(\Lambda)|\ll 1$ due to   symmetries of the new-physics sector.
For example, if the Higgs is a pseudo-Goldstone boson arising from a new  strong-sector,
$\kappa_{BB}(\Lambda)$ is protected by a shift symmetry and can only be generated by  loops involving SM  couplings,
while  $\kappa_{HB}(\Lambda)=\kappa_{HW}(\Lambda)\sim g_H^2/(16\pi^2)$ if the strong sector has an accidental 
custodial O(4) symmetry~\footnote{We have O(4) $\simeq \rm SU(2)_L\times SU(2)_R \times P_{LR}$  under which $\rm P_{LR}$ interchange $\rm L\leftrightarrow R$. Under this $\rm P_{LR}$ we have $c_{HW}\leftrightarrow c_{HB}$. 
To make the transformation properties under this symmetry more manifest,  it is better to work with ${\cal O}_{WB}$,
 which is even under $\rm P_{LR}$,
instead of  ${\cal O}_{BB}$.}
   \cite{Giudice:2007fh}. 
In this case \eq{final} could give the main correction
to the SM decay $h\rightarrow\gamma\gamma$ and could be as large as 
$\Delta\Gamma_{\gamma\gamma}/\Gamma^{\rm SM}_{\gamma\gamma}\sim g^2 v^2/\Lambda^2\log(\Lambda/m_h)$
if $g_H\sim 4\pi$.
Notice also that there can  be  finite one-loop corrections  to $\kappa_{\gamma\gamma}(m_h)$
from the operators (\ref{first6dim}) and (\ref{second6dim})
which can dominate over those in \eq{final}.
These were  calculated in ref.~\cite{Giudice:2007fh}.


A similar analysis can be performed for $ \kappa_{\gamma \widetilde\gamma}$, with the simplification that the operator identities corresponding to Eqs.~(\ref{OpId1}}) and (\ref{OpId2}) are, for the dual field strengths: 
 \begin{eqnarray}
&&
{\cal O}_{H\widetilde B}+\frac{1}{4}{\cal O}_{W\widetilde B}+\frac{1}{4}{\cal O}_{B\widetilde B}\ \label{OpId3}  = 0\ ,\\
&&{\cal O}_{H\widetilde W}+\frac{1}{4}{\cal O}_{W\widetilde W}+\frac{1}{4}{\cal O}_{W\widetilde B} =  0\ ,
\label{OpId4}
\end{eqnarray}
due to the Bianchi identity.
The above equations do not mix tree and loop generated operators; hence, from the calculation of \cite{GJMT} with the set  
$\{\mathcal{O}_{B\widetilde B},\mathcal{O}_{W\widetilde W},\mathcal{O}_{W\widetilde B}\}$ one can obtain the $\gamma_{\gamma\widetilde\gamma}$ in terms of the coeficients of the operators $\{\mathcal{O}_{B\widetilde B},\mathcal{O}_{H\widetilde B},\mathcal{O}_{H\widetilde W}\}$ of our basis. One arrives at the expected result: $\gamma_{\gamma\widetilde\gamma}=d\kappa_{\gamma\widetilde\gamma}/d\log\mu$ is given by the same expression as $\gamma_{\gamma \gamma}$ but with the corresponding CP-odd coefficients instead of the CP-even ones.

\section{RGEs for 
$\bma{\kappa_{\gamma  Z}}$ and $\bma{ \kappa_{\gamma\widetilde Z} }$ and a new basis 
\label{sec:rencgZ}}

If we try to obtain the RGE for $\kappa_{\gamma Z}$ in the same way as for $\kappa_{\gamma \gamma}$,
we face the complication that $\kappa_{\gamma Z}$ depends not only on
$c'_{BB}$, $c'_{WW}$ and $c'_{WB}$, but also on $c'_{HB}$ and $c'_{HW}$, and these coefficients were not included in the calculation presented in ref.~\cite{GJMT}. In other words, one would need
to calculate the anomalous-dimension matrix elements $b'_{i,j}$ for $i=\{HW, HB\}$
and $j=\{WW,WB,BB\}$, or, in our basis, to complete the $3\times 3$
anomalous-dimension matrix for $\kappa_{BB}, \kappa_{HW}, \kappa_{HB}$.  

\begin{figure}[t]
$$\includegraphics[width=0.3\textwidth]{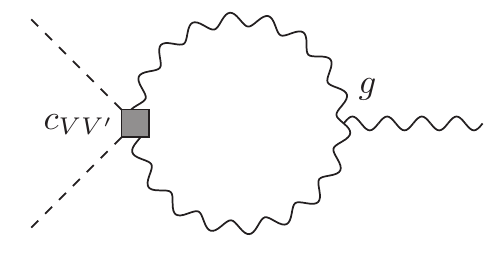}    \qquad
\includegraphics[width=0.3\textwidth]{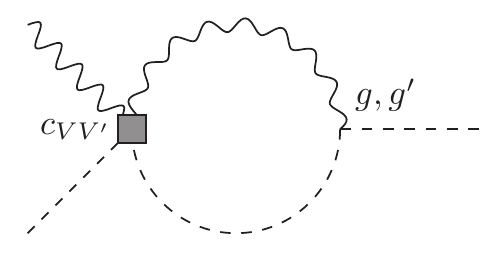}$$
\begin{center}\parbox{14cm}{
\caption{\emph{The only two diagrams that could give a contribution (at one loop) from ${\cal O}_{WW}$, ${\cal O}_{BB}$ and ${\cal O}_{WB}$
(with coefficient generically denoted as $c_{VV'}$ in the figure) to the renormalization of ${\cal O}_{HW}$ and ${\cal O}_{HB}$ (or to ${\cal O}_W$ and ${\cal O}_B$).}}}
 \end{center}
\end{figure}

We can circumvent this difficulty by realizing that the operators ${\cal O}_{WW}, {\cal O}_{BB}$ and ${\cal O}_{WB}$ do not enter in the (one-loop) RGEs for $c'_{HW}$ and $c'_{HB}$, so that
the matrix elements required to get $\gamma_{\gamma Z}$ are in fact zero. In order to see this, notice that both ${\cal O}_{HW}$ and ${\cal O}_{HB}$ include the trilinear pieces (with two Higgses and one gauge boson):
\bea
{\cal O}_{HW} &=&2ig(\partial^\mu H)^\dagger \sigma^a (\partial^\nu H)\partial_\mu W^a_\nu+\cdots\ ,\nonumber\\
{\cal O}_{HB} &= &2ig'(\partial^\mu H)^\dagger  (\partial^\nu H)\partial_\mu B_\nu+\cdots\ , 
\label{trilinears}
\eea
while ${\cal O}_{WW}, {\cal O}_{BB}$ and ${\cal O}_{WB}$ have two Higgses and at least two gauge bosons. Therefore,
in order to generate (at one loop) trilinears like those in (\ref{trilinears}), the only possibility is that one of the two gauge boson legs is attached to the other gauge boson leg or to one of the Higgs legs (see figure~1). In the first case (fig.~1, left diagram) it is clear that the resulting Higgs structure for the operator generated is
either $|H|^2$ or $H^\dagger \sigma^a H$ and not that in (\ref{trilinears}) (in fact, the diagram is zero). In the second case (fig.~1, right diagram)  the only structures that result
are either $\partial^\mu H^\dagger \partial^\nu(H B_{\mu\nu})$
or  $\partial^\mu H^\dagger \sigma^a\partial^\nu(H W^a_{\mu\nu})$, which give zero after integrating by parts.

We can therefore extract $\gamma_{\gamma Z}$ following the same procedure used for $\gamma_{\gamma\gamma}$ in the previous section, and we obtain
\be
16\pi^2\gamma_{\gamma Z}= \kappa_{\gamma Z} \left[6y_t^2+12\lambda-\frac{7}{2} g^2-\frac{1}{2}{g'}^2\right]
+(\kappa_{HW}+\kappa_{HB}) \left[2 g^2-3 e^2-2 \lambda  \cos (2 \theta_w)\right]\ ,
\ee
and a similar expression for  $\gamma_{\gamma\widetilde  Z}$ with the corresponding CP-odd operator coefficients instead of the CP-even ones.

The arguments we have used to prove that ${\cal O}_{WW}, {\cal O}_{BB}$ and ${\cal O}_{WB}$  do not enter into the anomalous dimensions of ${\cal O}_{HW}$ and ${\cal O}_{HB}$ can be applied in exactly the same way to prove that they do not generate radiatively the operators ${\cal O}_{W}$ and ${\cal O}_{B}$ which have exactly the same  trilinear structures displayed in \eq{trilinears} for ${\cal O}_{HW}$ and ${\cal O}_{HB}$. 
This immediately implies that the $5\times 5$ matrix of anomalous dimensions will be block diagonal if instead of using the bases in (\ref{B1}) and (\ref{B2}), we use instead the basis
\be
B_3 =\{{\cal O}_{BB}, {\cal O}_{WW},{\cal O}_{WB},{\cal O}_{W}, {\cal O}_{B}\}\ .
\label{B3}
\ee
Calling $\hat c_i, \hat\kappa_i$ the operator coefficients in this basis, we have
\be
\frac{d}{d\log\mu}
\left(
\begin{array}{c}
\hat \kappa_{BB}\\
\hat \kappa_{WW}\\
\hat \kappa_{WB}\\
\hat c_W\\
\hat c_B
\end{array}
\right)
=
\left(
\begin{array}{cc}
\hat\Gamma & 0_{3\times 2}\\
0_{2\times 3}& \hat X
\end{array}
\right)
\left(
\begin{array}{c}
\hat \kappa_{BB}\\
\hat \kappa_{WW}\\
\hat \kappa_{WB}\\
\hat c_W\\
\hat c_B
\end{array}
\right)\ .
\label{Gamma3}
\ee

Taking the anomalous-dimension matrix in the simple form (\ref{Gamma3}) as starting point, it is a trivial exercise  to transform it to other bases. 
In the GJMT basis one gets
\be
\frac{d}{d\log\mu}
\left(
\begin{array}{c}
c'_{BB}\\
c'_{WW}\\
c'_{WB}\\
c'_{HW}\\
c'_{HB}
\end{array}
\right)
=
\left(
\begin{array}{cc}
\hat \Gamma & Y'\\
0_{2\times 3}& \hat X
\end{array}
\right)
\left(
\begin{array}{c}
c'_{BB}\\
c'_{WW}\\
c'_{WB}\\
c'_{HW}\\
c'_{HB}
\end{array}
\right)\ .
\label{Gamma2}
\ee
The $3\times 3$ upper-left block is therefore given by the expression calculated in \cite{GJMT}:
\be
\hat \Gamma=\frac{1}{16\pi^2}
\left(
\begin{array}{ccc}
6y_t^2+12\lambda-\frac{9}{2}g^2+\frac{1}{2}{g'}^2&0&3g^2\\
0&6y_t^2+12\lambda-\frac{5}{2}g^2-\frac{3}{2}{g'}^2&{g'}^2\\
2{g'}^2&2g^2&6y_t^2+4\lambda+\frac{9}{2}g^2-\frac{1}{2}{g'}^2\\
\end{array}
\right)\ ,
\label{Gamma3UL}
\ee
while the $2\times 2$ lower-right block $\hat X$ has not been fully calculated in the literature.
This lack of knowledge affects also the ${3\times 2}$ block $Y'$, which depends on the entries of $\hat X$.

In basis $B_1$ one gets instead:
\be
\frac{d}{d\log\mu}
\left(
\begin{array}{c}
\kappa_{BB}\\
\kappa_{HW}\\
\kappa_{HB}\\
c_{W}\\
c_{B}
\end{array}
\right)
=
\left(
\begin{array}{cc}
\Gamma & 0_{3\times 2}\\
Y& \hat X
\end{array}
\right)
\left(
\begin{array}{c}
\kappa_{BB}\\
\kappa_{HW}\\
\kappa_{HB}\\
c_{W}\\
c_{B}
\end{array}
\right)\ ,
\label{Gamma1}
\ee
where now
\be
\Gamma=\frac{1}{16\pi^2}
\left(
\begin{array}{ccc}
6y_t^2+12\lambda-\frac{9}{2}g^2-\frac{3}{2}{g'}^2&\frac{3}{2}g^2-2\lambda&\frac{3}{2}g^2-2\lambda\\
0&6y_t^2+12\lambda-\frac{5}{2}g^2-\frac{1}{2}{g'}^2&{g'}^2\\
-8{g'}^2&9g^2-8\lambda&6y_t^2+4\lambda+\frac{9}{2}g^2+\frac{1}{2}{g'}^2\\
\end{array}
\right)\ ,
\label{Gamma1UL}
\ee
while $Y$ is also dependent on the unknown coefficients of $\hat X.$\footnote{Note that the lower-right block $\hat X$ is exactly the same in all the three bases considered.}
 We can  reexpress $\Gamma$
 in terms of the physically relevant combinations of coefficients $\kappa_{\gamma\gamma}$ and  $\kappa_{\gamma Z}$ defined  in (\ref{cggcgz}) plus the orthogonal combination $\kappa_{ort}\equiv \kappa_{HW}+\kappa_{HB}$. One gets
\be
\frac{d}{d\log\mu}
\left(
\begin{array}{c}
\kappa_{\gamma\gamma}\\
\kappa_{\gamma Z}\\
\kappa_{ort}
\end{array}
\right)
=
\Gamma_{o}
\left(
\begin{array}{c}
\kappa_{\gamma\gamma}\\
\kappa_{\gamma Z}\\
\kappa_{ort}
\end{array}
\right)\ ,
\label{Gammaph}
\ee
where
\be
\Gamma_{o}=\frac{1}{16\pi^2}
\left(
\begin{array}{ccc}
6y_t^2+12\lambda-\frac{9}{2}g^2-\frac{3}{2}{g'}^2&0&\frac{3}{2}g^2-2\lambda\\
0&6y_t^2+12\lambda-\frac{7}{2}g^2-\frac{1}{2}{g'}^2&2 g^2-3 e^2-2 \lambda  \cos (2 \theta_w)\\
-16e^2&-4g^2+4{g'}^2&6y_t^2+4\lambda+\frac{11}{2}g^2+\frac{1}{2}{g'}^2\\
\end{array}
\right)\ ,
\label{Gamma1UL}
\ee
from which we explicitly see that $\kappa_{\gamma Z}$ does not renormalize $\kappa_{\gamma \gamma}$ and vice versa.

We have seen that the expression for the anomalous-dimension matrix takes the simplest block-diagonal form in basis $B_3$. 
This basis has also the virtue of $B_1$ of keeping separated current-current operators from one-loop suppressed ones.
Indeed, using Eqs.~(\ref{OpId1}) and (\ref{OpId2}), we can reach $B_3$  from $B_1$ by trading two one-loop suppressed operators,
  ${\cal O}_{HW}$ and ${\cal O}_{HB}$,
  by
other two  one-loop suppressed ones,  ${\cal O}_{WW}$ and ${\cal O}_{WB}$.
In spite of the fact that the anomalous-dimension matrix gets its
simplest form in basis $B_3$,  there are other advantages in using basis $B_1$.
For example, in $B_1$ only one operator  contributes to $h\rightarrow \gamma\gamma$, while there are three in basis $B_3$.
Also  $B_1$  is a more suitable basis to describe the low-energy effective theory expected for a pseudo-Goldstone Higgs boson \cite{Giudice:2007fh}, as it clearly identifies operators invariant under constant shifts $H\rightarrow H+ c$.

\section{Dipole operators}

The above analysis  can be easily extended to include  contributions from operators involving SM fermions.
We will limit the discussion here to the up-quark sector, having in mind  possible large contributions from the top. 
The extension to other SM fermions is straightforward.
We organize again the operators as tree-level and one-loop suppressed  ones.
Among the first type we have the operators already given in   \eq{oy}, \eq{first6dimF},  apart from 
four-fermion operators. 
In Section~\ref{section3}, however, 
we   already showed  that they cannot contribute to 
 the anomalous dimension  of the  operators (\ref{third6dim1})-(\ref{third6dimCP3}) at the one-loop level.
 Among  one-loop suppressed operators made with SM fermions,
we have the   dipole operators
\bea
{\cal O}_{DB} &=&y_u\bar Q_L \sigma^{\mu \nu} u_R\, \widetilde H g'B_{\mu \nu}\ , \nonumber\\
{\cal O}_{DW}&=&y_u
 \bar Q_L \sigma^{\mu \nu} u_R\,  \sigma^a\widetilde HgW^a_{\mu \nu}\ , \nonumber\\
{\cal O}_{DG}&=&y_u
 \bar Q_L \sigma^{\mu \nu} T^a u_R\, \widetilde H g_sG^a_{\mu \nu}\, ,
\label{third6dimF}
\eea
where $T^a$ are the $SU(3)_C$ generators.
These operators can, in principle, give  contributions to
 other one-loop suppressed operators, as those relevant for $h\rightarrow\gamma\gamma,\gamma Z$.
We have calculated that, indeed, such contributions are nonzero:
\bea
16\pi^2 \gamma_{\gamma\gamma} &=& 8y_u^2 N_cQ_u{\rm Re}[\kappa_{DB}+\kappa_{DW}] \ , \nonumber\\
16\pi^2\gamma_{\gamma\widetilde \gamma} &=& -8y_u^2N_c  Q_u {\rm Im}[\kappa_{DB}+\kappa_{DW}]\ ,\nonumber\\ 
16\pi^2 \gamma_{\gamma Z} &=&4 y_u^2N_c\left\{\left(\frac{1}{2}-4Q_us_w^2\right){\rm Re}[\kappa_{DB}]+\left(\frac{1}{2}+2Q_uc_{2w}\right){\rm Re}[\kappa_{DW}]\right\}\ , \nonumber\\
16\pi^2\gamma_{\gamma\widetilde Z} &=& -4 y_u^2N_c\left\{\left(\frac{1}{2}-4Q_us_w^2\right){\rm Im}[\kappa_{DB}]+\left(\frac{1}{2}+2Q_uc_{2w}\right){\rm Im}[\kappa_{DW}]\right\}\, ,
\label{rgdp}
\eea
where $N_c=3$, $Q_u=2/3$ is the electric charge of the up-quark, $c_{2w}=\cos(2\theta_w)$,
 and the $\kappa_i$ are the one-loop suppressed coefficients of the operators of \eq{third6dimF}, {\it i.e.} $\delta{\cal L}= \kappa_i {\cal O}_i/\Lambda^2+{\rm h.c.}$.
In the $B_3$ basis, \eq{rgdp}  arises from 
\be
\frac{d}{d\log\mu}
\left(
\begin{array}{c}
\hat \kappa_{BB}\\
\hat \kappa_{WW}\\
\hat \kappa_{WB}
\end{array}
\right)
=\frac{4N_cy_u^2}{16\pi^2}
\left(
\begin{array}{ccc}
  0& Y^u_{L}+Y^u_{R}\\
 1/2& 0\\
-( Y^u_{L} + Y^u_{R})& -1/2
\end{array}
\right)
\left(
\begin{array}{c}
\hat \kappa_{DW}\\
\hat \kappa_{DB}
\end{array}
\right)\ ,
\ee
where $Y^u_L=1/6$ and $Y^u_R=2/3$ are the up-quark hypercharges.
Similar results follow for the RGE of the Higgs couplings to gluons,  $\kappa_{GG}$ and $ \kappa_{G\widetilde G}$~\footnote{This contradicts the results of ref.~\cite{cGG}, which finds a  cancelation of the logarithmic divergence responsible for the non-zero $\gamma_{GG}$. 
A similar cancelation found in \cite{cggdipole} has  been however recently corrected, as    
C. Grojean and G. Servant have pointed out to us.  
}
\be
16\pi^2 \gamma_{GG} = 4y_u^2 {\rm Re}[\kappa_{DG}] \ , \quad
16\pi^2\gamma_{G\widetilde G} =-4y_u^2  {\rm Im}[\kappa_{DG}]\, .
\ee

\section{The ${\bma S}$ parameter
\label{sec:Sparam}}

As we have shown above,
   the Wilson coefficients of the current-current operators (\ref{first6dim})-(\ref{second6dim}) 
do not enter in the one-loop RGEs  of   the $\kappa_i$,
but only  in their own  RGEs.
In particular,  the only   operators with two Higgs bosons and gauge bosons
affected   by  $c_{H,T}$ at  one loop 
are ${\cal O}_W$ and  ${\cal O}_{B}$ and not those relevant for $h\rightarrow \gamma\gamma,\gamma Z$.
Indeed,  an explicit calculation gives
\be
\gamma_W=\frac{dc_{W}}{d\log\mu}  =-\frac{g_H^2}{16\pi^2}\ \frac{1}{3} (c_H  +c_T) \ ,\quad
\gamma_B=\frac{dc_{B}}{d\log\mu}  =     -\frac{g_H^2}{16\pi^2}\
 \frac{1}{3} (c_H   +5 c_T )\, .
\label{betaWB}
\ee
In  the  basis $B_1$ of Section~\ref{Dimension-six Operator Basis}, these are the only two Wilson coefficients that enter 
in the  $S$-parameter \cite{Peskin:1991sw}.
We have  $S=4\pi v^2 [c_W(m_Z)+c_B(m_Z)]/\Lambda^2$
where $c_{W,B}(m_Z)$ is the value of the coefficient at the $Z$ mass.
The contributions from \eq{betaWB} to $c_{W,B}(m_Z)$ can be sizeable for $g_H\gg 1$ \cite{Barbieri:2007bh}, although the value of $c_T$
is highly constrained from the $T$-parameter \cite{Giudice:2007fh}.
The anomalous dimensions $\gamma_{W}$  and $\gamma_B$ can also receive corrections proportional to $c_{W,B}$, or from 
  one-loop suppressed operators, such as ${\cal O}_{BB}$.
Nevertheless these contributions are not expected 
   to be sizeable. The coefficients $c_{W}$ and $c_{B}$
   already   contribute at  tree-level to $S$, while 
the contributions to $S$ from $\kappa_{i}$
are  expected to be small,   $\delta \gamma_W=O(\kappa_{i}/(16\pi^2))$. 
Notice that  basis $B_1$ makes very clear the separation between the relevant contributions to $S$ that come from tree-level operators  and those to $\kappa_{\gamma\gamma}$, which are from  one-loop suppressed operators.

In the GJMT basis  the contribution to $S$ arises from
 the operator ${\cal O}_{WB}$ and one has  
 $S=16\pi v^2 c'_{WB}(m_Z)/\Lambda^2$.
In ref.~\cite{GJMT}, a partial calculation
of the anomalous dimension of ${\cal O}_{WB}$ was given. 
Nevertheless,  if the interest
is to calculate   the running of $c'_{WB}$ 
in universal theories   in which  $c_{W}$ and $c_B$ encode the dominant effects [apart from $c_{H,T}$ whose effects are given in \eq{betaWB}], one also needs,
as \eq{invdict} shows,  to include the effects of $c'_{HW}$ and  $c'_{HB}$ given in ref.~\cite{zeppe}.
This is again due to the fact that the GJMT basis  mixes current-current operators with one-loop suppressed ones.

Finally, let us comment on the relation between our basis and one of  the most used in the literature,
 the one originally given in ref.~\cite{Buchmuller:1985jz}.
After eliminating  redundant operators, one ends up with 59 independent operators as listed  in ref.~\cite{Grzadkowski:2010es}.
This basis also keeps separate tree-level operators 
from one-loop suppressed ones.
The set of one-loop suppressed  operators is  different from ours though:  they use
$\{{\cal O}_{WW}, {\cal O}_{WB},{\cal O}_{W\widetilde W}, {\cal O}_{W\widetilde B}\}$ instead of  our
$\{{\cal O}_{H W}, {\cal O}_{H B},{\cal O}_{H\widetilde W}, {\cal O}_{H\widetilde B}\}$. The change of basis is given in 
Eqs.~(\ref{OpId1}), (\ref{OpId2}), (\ref{OpId3}) and  (\ref{OpId4}).
For the tree-level operators they use the minimal set of 3 operators made of  SM bosons, in particular ${\cal O}_H$, ${\cal O}_T$ and ${\cal O}_6$,
while the rest of operators involves SM fermions:  those given in  \eq{oy}, \eq{first6dimF}   and 
four-fermion operators.
As explained in the Appendix, we can reach this set of operators from our basis  by  performing field redefinitions.
The basis of refs.~\cite{Buchmuller:1985jz,Grzadkowski:2010es} is, however, not very convenient 
for parametrizing the effects of universal theories.
Although only a few operators  parametrize these theories in our basis  (see Section~\ref{Dimension-six Operator Basis}),
 in the basis of
refs.~\cite{Buchmuller:1985jz,Grzadkowski:2010es} they require
 a  much larger set  of operators.
  In particular, the two tree-level operators ${\cal O}_W$ and 
${\cal O}_B$   are written  in the basis of
refs.~\cite{Buchmuller:1985jz,Grzadkowski:2010es}  as
\bea
c_W{\cal O}_W&\rightarrow&  c_W\frac{g^2}{g^2_H}\left[-\frac{3}{2}{\cal O}_H+2{\cal O}_6+\frac{1}{2}{\cal O}_y+  \frac{1}{4}\sum_f{\cal O}_L^{f\, (3)}\right]\, ,\nonumber\\
c_B{\cal O}_B&\rightarrow&  c_B\frac{g^{\prime\, 2}}{g^2_H}\left[-\frac{1}{2}{\cal O}_T+\frac{1}{2}\sum_f\left(Y_L^f {\cal O}^f_L+ Y_R^f{\cal O}^f_R\right)\right]\, ,
\label{tobg}
\eea
where $Y_L^f$ and $Y_R^f$ are the hypercharges of the left and right handed fermions, respectively. We can see from (\ref{tobg}) that   the Wilson coefficients in the basis of \cite{Buchmuller:1985jz,Grzadkowski:2010es}
 are correlated, so that one should include them all in operator analyses of universal theories.
As far as the anomalous-dimension matrix is concerned, the basis 
of \cite{Buchmuller:1985jz,Grzadkowski:2010es} keeps also the same block-diagonal form as the basis of $B_3$,
since loop-suppressed operators 
$\{{\cal O}_{BB},{\cal O}_{WW}, {\cal O}_{WB},{\cal O}_{B\widetilde B}, {\cal O}_{W\widetilde W}, {\cal O}_{W\widetilde B}\}$
do not mix with current-current ones.


\section{Conclusions}

After the recent discovery of the Higgs boson at the LHC, it is natural to start  precision studies of the Higgs couplings to SM particles. The
$h \rightarrow \gamma \gamma$ decay is of special importance because of its clean experimental 
signature, and also because its measurement  hints at a possible discrepancy with the SM prediction \cite{Higgs}.
In this article we have  analyzed potential effects of new physics in this decay rate (together with the closely related one $h\rightarrow \gamma Z$) following the effective
Lagrangian approach, where one enlarges the SM Lagrangian with a set of dimension-six operators.
The  choice of the  operator basis  has been crucial to 
make the calculations
 simple and  transparent.
 We have shown  the convenience of working in bases that classify operators in two groups.
The first  is formed by operators which can arise
from  tree-level  exchange of heavy states  under the assumption of minimal coupling.
This group contains  operators
that can be written as  a product of local currents.
A second group contains operators   that are generated,
from  weakly-coupled renormalizable theories,
at the loop-level, and thus have suppressed coefficients.
Following this criteria, we have defined our basis in  \eq{6dim},
where  we have symbolized the Wilson coefficients of the operators of the first group by 
$c_{i_1}$ and $c_{i_2}$, while the  Wilson  coefficients of the  second group, which contain a loop factor, 
have been written as   $\kappa_{i_3}$. 

The operators relevant for 
$h \rightarrow \gamma \gamma,\gamma Z$
are, as  expected, of the second group,
specifically  ${\cal O}_{BB}$, ${\cal O}_{HW}$  and ${\cal O}_{HB}$
and their CP-odd counterparts. 
We have been interested in  the anomalous dimensions of these operators that can be
generically written as
\begin{equation}
16 \pi^2 \frac{d \kappa_{j_3}}{d \log \mu} = \sum_{i_1} b_{j_3,i_1} c_{i_1} + \sum_{i_2} b_{j_3,i_2} c_{i_2}  
+ \sum_{i_3} b_{j_3,i_3} \kappa_{i_3}\, , 
\end{equation}
where $j_3=BB,HW,HB,B\widetilde B,H\widetilde W, H\widetilde B$.
The main purpose of this article has been 
 to calculate $b_{j_3,i_1} $ and $b_{j_3,i_2}$. 
 Since the corresponding coefficients $c_{i_1}$ and  $c_{i_2}$ can be of order one,
   the RG evolution  can 
 enhance the new-physics effect  on  $\kappa_{i_3}$ by a factor $\log(\Lambda/m_h)$. 
Our main result is that   such enhancement is not present, 
because the corresponding elements of the anomalous-dimension matrix
vanish
\begin{equation}
b_{j_3,i_1} = b_{j_3,i_2} = 0 \, .
\end{equation}
Therefore, tree-level (current-current) operators do not contribute to the RGEs of the   one-loop suppressed operators relevant for the $\gamma \gamma$ and $\gamma Z$ Higgs  decay.
This differs from ref.~\cite{GJMT}, which claims that such enhancement
 exists.  
Nevertheless, we  have shown that the results of ref.~\cite{GJMT}  can be put in agreement  with our result
when one takes into account all operators in their basis.
The  anomalous-dimension matrix elements 
$b_{j_3,i_3}$ are however nonzero.
Using ref.~\cite{GJMT},
we have  been able to calculate these elements
for the case of $\kappa_{BB}$ relevant for $h\rightarrow\gamma\gamma$.
The result is given in \eq{final} (and its CP-odd analog).

We have also obtained the RGEs for $\kappa_{HW}$ and $\kappa_{HB}$, \eq{Gamma1},
which affect  the decay $h\rightarrow \gamma Z$, by realizing that the operators ${\cal O}_{BB}$, ${\cal O}_{WW}$, ${\cal O}_{WB}$ (used in \cite{GJMT}) do not renormalize (at one-loop) ${\cal O}_{HW}$, ${\cal O}_{HB}$ (nor ${\cal O}_W$, ${\cal O}_B$). Exploiting this fact, we have further clarified the structure of the anomalous-dimension matrix for these operators, showing that it takes a particularly simple block-diagonal form in the basis $B_3$
of Eq.~(\ref{B3}).  The  tree-level operators ${\cal O}_B$ and ${\cal O}_W$ do not mix with the one-loop operators ${\cal O}_{WW}$, ${\cal O}_{BB}$, ${\cal O}_{WB}$ and vice versa, as  Eq.~(\ref{Gamma3}) shows. 
Enlarging this basis with dipole-moment operators for the SM fermions, we have further computed the effect of such dipoles on $h\rightarrow \gamma\gamma,\gamma Z$.

To conclude, we have discussed how the appropriate choice of operator basis can shed light on the physical structure behind the renormalization mixing of operators and reveal hidden simplicities in the structure of the matrix of anomalous
dimensions that describes such mixing. 

\section*{Acknowledgements}
We thank  S. Gupta, M. Jamin, D. Marzocca, G. Servant, M. Trott and especially C. Grojean for useful discussions.
J.R.E. thanks CERN for hospitality and partial financial support during the final stages of this work.
This work has been partly supported
by Spanish Consolider Ingenio 2010 Programme CPAN (CSD2007-00042) and
the Spanish Ministry MICNN under grants FPA2010-17747 and
FPA2011-25948; and the Generalitat de Catalunya grant 2009SGR894.
The work of A.P. has also been  supported by  the ICREA Academia Program. The work of J.E.M. has been supported by the Spanish Ministry MECD through the FPU grant AP2010-3193.

\newpage
\section*{Appendix: Change of basis by field redefinitions}

The following field redefinitions
\bea
H\rightarrow H \left(1+\alpha_1 g_H^2|H|^2/\Lambda^2\right)\ ,&&
H\rightarrow H\left(1-\alpha_2 g_H^2 m^2/
\Lambda^2\right)+\alpha_2 g_H^2 (D^2 H)/\Lambda^2\ ,\nonumber\\
B_\mu \rightarrow B_\mu + i g'\alpha_B  (H^\dagger
\lra{D^\mu} H)/\Lambda^2\ ,&&
W^a_\mu \rightarrow W^a_\mu + i g \alpha_W  (H^\dagger\sigma^a
\lra{D^\mu} H)/\Lambda^2\ ,\nonumber\\
B_\mu \rightarrow B_\mu + \alpha_{2B}  (\partial^\nu B_{\nu\mu})/\Lambda^2\ ,&&
W^a_\mu \rightarrow W^a_\mu + \alpha_{2W}  (D^\nu W^a_{\nu\mu})/\Lambda^2\ ,
\label{fieldredef}
\eea
where the $\alpha_i$ are arbitrary parameters,
induce the following shifts in the coefficients of the dimension-six  
operators of  Eqs.~(\ref{first6dim}) and (\ref{second6dim}) plus ${\cal O}_{4K}=|D_\mu^2 H|^2$:~\footnote{Shifts of 
order $m^2/\Lambda^2$ are also induced on the renormalizable dimension-4 SM operators.}
\bea
&&c_H\rightarrow c_H +2(\alpha_1+2\lambda\alpha_2)-\alpha_W g^2/g_H^2\ ,\nonumber\\
&&c_r\rightarrow c_r +2(\alpha_1+2\lambda\alpha_2)+2\alpha_W g^2/g_H^2\ ,\nonumber\\
&&c_6\rightarrow c_6-4\alpha_1\ ,\nonumber\\
&&c_T\rightarrow c_T- \alpha_B {g'}^2/g_H^2\ ,\nonumber\\
&&c_B\rightarrow c_B-2\alpha_B-\alpha_{2B}\ ,\nonumber\\
&&c_W\rightarrow c_W-2\alpha_W-\alpha_{2W}\ ,\nonumber\\
&&c_{2W}\rightarrow c_{2W}-2\alpha_{2W}\ ,\nonumber\\
&&c_{2B}\rightarrow c_{2B}-2\alpha_{2B}\ ,\nonumber\\
&&c_{K4}\rightarrow c_{K4}-2\alpha_2g_H^2\, .
\eea
Notice that only operators of tree-level type are shifted. This is not a coincidence: diagrammatically, 
a field redefinition $\Phi\rightarrow \Phi + J[\phi_i,\phi_j,...]$
(with $J$ some current with the same quantum numbers as $\Phi$ and
dependent on some other fields $\phi_i$) corresponds to a $\Phi$ leg splitting in several $\phi_{i,j}...$ legs. Then, an operator generated by such field redefinition corresponds to a {\it tree-level} diagram with a heavy state of mass $\sim \Lambda$ (with the same quantum numbers of $\Phi$) as an internal propagator.

Using this shift freedom, we can trade 6 out of the 9 tree-level  operators listed in section~\ref{Dimension-six Operator Basis} (${\cal O}_{2G}$ is irrelevant for our discussion)
and leave only ${\cal O}_H$, ${\cal O}_T$ and ${\cal O}_6$  plus operators made of fermions: those in (\ref{oy}), (\ref{first6dimF}) and four-fermion operators.
The shift parameters are arbitrary, and therefore physical quantities 
can only depend on the three following shift-invariant combinations
(we reserve capital letters for such physical combinations of coefficients):
\bea
C_H&\equiv & c_H - c_r -\frac{3g^2}{4g_H^2}(2c_W-c_{2W})\ ,\nonumber\\
C_T&\equiv & c_T-\frac{{g'}^2}{4 g_H^2}(2c_B-c_{2B})\ ,\nonumber\\
C_6&\equiv & c_6 +2c_r+\frac{g^2}{g_H^2}(2c_W-c_{2W})+4 \frac{\lambda}{g_H^2}c_{K4}\ .
\label{physcoef}
\eea

One concern in analyzing operator renormalization (for instance if one is interested in calculating the renormalization group equations for the $c_i$ Wilson coefficients) is that the
redundant operators we have decided to remove from the Lagrangian
might be generated radiatively anyway. The simplest way to deal with that complication is to write RGEs for the $C_i$'s, the physical combinations of coefficients, which must only depend on the $C_i$'s themselves. In those equations one can then consistently set equal to zero the coefficients of the redundant operators appearing implicitly in the
$C_i$'s.  In our particular example, this means that the RGEs of all
our tree-level operators can be reduced to a $3\times 3$ anomalous-dimension matrix for $C_H$, $C_T$ and $C_6$. For this reason, the main question  discussed in this paper  about  the possible mixing of tree-level operators with loop-induced ones
through their RGEs, reduces to the question of whether ${\cal O}_H$,
${\cal O}_T$ and ${\cal O}_6$ do mix with them.

The field redefinitions listed in Eq.~(\ref{fieldredef}) also induce shifts of the coefficients of  dimension-six operators that involve fermions. In addition, further field redefinitions of fermions themselves [like
$f_{L,R}\rightarrow f_{L,R} (1+\alpha_{f_{L,R}} |H|^2/\Lambda^2)$ or $B_\mu \rightarrow 
B_\mu + \sum_{f}\alpha^B_{f_{L,R}} (f_{L,R}\gamma_\mu f_{L,R})/\Lambda^2$, etc.] can be used
in the same way to remove many of these fermionic operators.  Besides 4-fermion operators, the operators involving only fermions plus gauge bosons can be eliminated completely by such shifts and the list of dimension-six operators with Higgs and fermions can be reduced to  operators of the type ${\cal O}_y$, ${\cal O}_L^f$, ${\cal O}_R^f$ and ${\cal O}_L^{f\, (3)}$.



\begin{thebibliography}{99}
%
\bibitem{Higgs} 
  S.~Chatrchyan {\it et al.}  [CMS Collaboration],
  Phys.\ Lett.\ B {\bf 710} (2012) 403
  [hep-ex/1202.1487];
  G.~Aad {\it et al.}  [ATLAS Collaboration],
  Phys.\ Rev.\ Lett.\  {\bf 108} (2012) 111803
  [hep-ex/1202.1414].

\bibitem{GJMT}
  C.~Grojean, E.~E.~Jenkins, A.~V.~Manohar and M.~Trott,
  [hep-ph/1301.2588].

\bibitem{Grinstein:1990tj}
  B.~Grinstein, R.~P.~Springer and M.~B.~Wise,
  Nucl.\ Phys.\ B {\bf 339} (1990) 269.



\bibitem{Giudice:2007fh}
  G.~F.~Giudice, C.~Grojean, A.~Pomarol and R.~Rattazzi,
  JHEP {\bf 0706} (2007) 045
  [hep-ph/0703164].


\bibitem{LRV}
  I.~Low, R.~Rattazzi and A.~Vichi,
  JHEP {\bf 1004} (2010) 126
  [hep-ph/0907.5413].
  



\bibitem{Shifman:1979eb}
  M.~A.~Shifman, A.~I.~Vainshtein, M.~B.~Voloshin and V.~I.~Zakharov,
  Sov.\ J.\ Nucl.\ Phys.\  {\bf 30} (1979) 711
   [Yad.\ Fiz.\  {\bf 30} (1979) 1368].


\bibitem{cGG}
  D.~Choudhury and P.~Saha,
  JHEP {\bf 1208} (2012) 144
  [hep-ph/1201.4130].

\bibitem{cggdipole}
  C.~Degrande, J.~M.~Gerard, C.~Grojean, F.~Maltoni and G.~Servant,
  JHEP {\bf 1207} (2012) 036
  [hep-ph/1205.1065].

\bibitem{Buchmuller:1985jz}
  W.~Buchm\"uller and D.~Wyler,
  Nucl.\ Phys.\ B {\bf 268} (1986) 621.
  
\bibitem{Grzadkowski:2010es}
  B.~Grzadkowski, M.~Iskrzynski, M.~Misiak and J.~Rosiek,
  JHEP {\bf 1010} (2010) 085
  [hep-ph/1008.4884].


\bibitem{Peskin:1991sw}
  M.~E.~Peskin and T.~Takeuchi,
  Phys.\ Rev.\ D {\bf 46} (1992) 381.

\bibitem{Barbieri:2007bh}
  R.~Barbieri, B.~Bellazzini, V.~S.~Rychkov and A.~Varagnolo,
  Phys.\ Rev.\ D {\bf 76} (2007) 115008
  [hep-ph/0706.0432].

\bibitem{zeppe} 
  K.~Hagiwara, S.~Ishihara, R.~Szalapski and D.~Zeppenfeld,
  Phys.\ Rev.\ D {\bf 48} (1993) 2182;
  K.~Hagiwara, R.~Szalapski and D.~Zeppenfeld,
  Phys.\ Lett.\ B {\bf 318} (1993) 155
  [hep-ph/9308347];
  S.~Alam, S.~Dawson and R.~Szalapski,
  Phys.\ Rev.\ D {\bf 57} (1998) 1577
  [hep-ph/9706542].
 


\end{thebibliography}
\end{document}